\documentclass[paper,notoc]{JHEP3} 

\JHEPspecialurl{http://jhep.sissa.it/JOURNAL/JHEP3.tar.gz}

\usepackage{amsmath,epsfig,multicol}

\newcommand{\be}{\begin{eqnarray}}
\newcommand{\ee}{\end{eqnarray}}
\newcommand{\nee}{\nonumber\end{eqnarray}}
\newcommand{\nn}{\nonumber}
\newcommand{\noi}{\noindent}

\newcommand{\sfrac}[2] {{\textstyle \frac{#1}{#2}}}
\newcommand{\smaf}[2]  {{\textstyle \frac{#1}{#2} }}

\newcommand{\eq}[1]  {\mbox{(\ref{eq:#1})}}
\newcommand{\fig}[1] {\mbox{Fig.~\ref{fig:#1}}}
\newcommand{\Fig}[1] {\mbox{Figure~\ref{fig:#1}}}

\def\b               {\beta}
\def\d               {\delta}
\def\g               {\gamma}
\def\G               {\Gamma}

\def\t               {\theta}

\def\x               {\chi}

\def\ti              {\tilde}

\def\snu             {\ti\nu}  
\def\stau            {\ti\tau}
\def\st              {\ti\nu}
\def\sb              {\ti\tau}
\def\ch              {\ti\x^\pm}

\def\nt              {\ti\x^0}

\def\sq              {\ti\ell}

\def\dcp             {\d^{CP}_{\tau\nu}}

\newcommand{\msnu}     {m_{\snu}}
\newcommand{\mstau}[1] {m_{\stau_{#1}}}
\newcommand{\mst}[1]   {\msnu}
\newcommand{\msb}[1]   {m_{\stau_{#1}}}
\newcommand{\mch}[1]   {m_{\ti \x^+_{#1}}}
\newcommand{\mnt}[1]   {m_{\ti \x^0_{#1}}}
\newcommand{\mhp}      {m_{H^+}}
\newcommand{\mh}       {m_{H^\pm}}

\def\PL              {P_L^{}}
\def\PR              {P_R^{}}

\def\Rsb             {R^{\,\sb}}

\def\Rsbs            {R^{\,\sb *}}

\def\fb              {${\rm fb}^{-1}$}
\def\gev             {{\rm GeV}}
\def\rzw             {\sqrt{2}}
\def\delr            {\!\stackrel{\leftrightarrow}{\partial^\mu}\!}

\renewcommand{\Re}{{{\cal R}e}}
\renewcommand{\Im}{{{\cal I}m}}

\def\lrd{\stackrel{\leftrightarrow}{\partial^{\mu}}}

\newcommand{\nutau}{
\nu_\tau\hspace{-4.8mm}
\raisebox{2.2mm}{\mbox{\tiny{(\raisebox{-0.4mm}{\footnotesize{$-$}})}}}}

\title{CP violation in charged Higgs boson decays\\
       into tau and neutrino}

\author{E. Christova\thanks{Visitor at CERN in October 2002.}\\
	Institute of Nuclear Research and Nuclear Energy, 
        Sofia 1784, Bulgaria\\
	E-mail: \email{echristo@inrne.bas.bg}}

\author{H. Eberl, W. Majerotto\\
	Institut f\"ur Hochenergiephysik der \"OAW, A-1050 Vienna, 
        Austria\\
	E-mail: \email{helmut@hephy.oeaw.ac.at, majer@hephy.oeaw.ac.at}}

\author{S. Kraml\\
	CERN, CH-1211 Geneva 23, Switzerland\\
	E-mail: \email{sabine.kraml@cern.ch}}

\preprint{CERN-TH/2002-309\\ HEPHY-PUB 767/02}

\abstract{We calculate the CP-violating rate asymmetry of $H^\pm$ 
decays into tau and neutrino at one loop in the MSSM with 
complex parameters. 
We find that the asymmetry is typically of the order of $10^{-3}$, 
depending mainly on the phases of the trilinear coupling $A_\tau$ 
and the gaugino mass $M_1$.}

\keywords{Supersymmetric Standard Model, Higgs Physics, CP violation}

\begin{document} 

\section{Introduction}

The CP violation observed in the kaon system and in $B$ meson decays 
appears to be consistent with the Standard Model (SM).
However, the baryon asymmetry in the universe requires a new source 
of CP violation beyond the CKM phase of the SM. Indeed, most extensions 
of the SM provide additional sources of CP violation. Searching for new 
effects of CP violation has become one of the most interesting ways to 
test physics beyond the SM~\cite{Ibrahim-Nath}.

Within the Minimal Supersymmetric Standard Model (MSSM)
with complex parameters, the new sources of CP violation  
are the phase of the higgsino mass parameter $\mu$, two phases of the 
gaugino masses $M_{i}, i = 1,2,3$, and the phases of the trilinear 
couplings $A_{f}$. Especially the latter ones are practically 
unconstrained.
Recently, we pointed out \cite{Christova:2002ke} that CP violation 
in the MSSM may lead to a difference in the partial decay widths of 
$H^+$ and $H^-$. 
More precisely, we showed that large phases of $A_{t}$, $A_{b}$ and/or 
$M_3$ can give a CP-violating asymmetry  
$\d^{CP}_{tb} = \left[\G\,(H^+\to t\bar{b}) - \G\,(H^{-}\to\bar{t}b)\right]/
                \left[\G\,(H^+\to t\bar{b}) + \G\,(H^{-}\to\bar{t}b)\right]$
of 10\,--\,15\% for $\mh > m_{\tilde{t}} + m_{\tilde{b}}$.

In this article, we consider the lepton decay channels of the charged 
Higgs bosons, $H^+\to\tau^+\nu_\tau$ and $H^-\to\tau^-\bar\nu_\tau$. 
In particular, we calculate the CP-violating asymmetry
\begin{equation}
  \dcp =
  \frac{\G\,(H^+\to\tau^+\nu_\tau)-\G\,(H^-\to\tau^-\bar\nu_\tau)}
       {\G\,(H^+\to\tau^+\nu_\tau)+\G\,(H^-\to\tau^-\bar\nu_\tau)}\,,
\label{eq:defDCP}
\end{equation}
at the one-loop level in the MSSM with explicit CP violation   
and discuss its parameter dependence. 
The decay into $\tau\nu$ may be important for relatively low masses 
of $H^\pm$ and large $\tan\beta$ as its branching ratio increases 
with increasing $\tan\beta$.  
For example, for 
$m_{H^+}$ = 250 GeV, we have 
$Br(H^+ \to t\bar b) = 0.90$ and
$Br(H^+ \to \tau^+\nu_\tau) = 0.06$ for $\tan\beta = 5$, and 
$Br(H^+ \to t\bar b) = 0.64$ and
$Br(H^+ \to \tau^+\nu_\tau) = 0.36$ for $\tan\beta = 30$. 
The asymmetry $\dcp$ is sensitive to the phases of the trilinear coupling $A_\tau$
and of the gaugino mass $M_1$. 
Although one expects $\dcp$ to be smaller than $\d^{CP}_{tb}$ 
due to the missing gluino exchange, it is an interesting quantity 
in the case  
$m_{\ti \tau} + m_{{\ti \nu}_\tau} < \mh < m_{\ti t} + m_{\ti b}$.

The article is organized as follows: in Sect.~2 we derive the basic 
formulae for the $H^\pm \to \tau\nu$ decay widths and define $\dcp$ 
in terms of CP-violating form factors $\d Y_\tau^{CP}$. 
In Sect.~3, we perform a numerical analysis. 
In Sect.~4, we summarize our results and comment on the 
measurability of $\dcp$. 
Appendices A and B contain the explicit formulae for the 
form factors, masses and couplings.

\section{Decay widths at tree level and one loop}

At lowest order, the widths of the $H^\pm \to \tau^\pm\,\,\nutau$ decays 
are given by
\begin{equation}
  \G^{\,LO}\,(H^\pm \to \tau\nu) = 
  \frac{(\mh^2 - m_\tau^2)^2\,y_\tau^2}{16\pi\mh^3}\,,
\end{equation}
where $y_\tau = h_\tau\sin\b$, $h_\tau$ being the tau Yukawa 
coupling, $h_\tau = g m_\tau/\left(\sqrt{2}\,m_W \cos\beta\right)$.  
At tree level, the amplitude is real and the decay widths are always 
equal, 
$\G^{\,LO}(H^+\to\tau^+\nu_\tau) = 
 \G^{\,LO}(H^-\to\tau^-\bar\nu_\tau)$. 
However, once loop corrections with complex couplings are included, 
we have $y_\tau\to Y_\tau^{\pm} = y_\tau^{} + \d Y_\tau^{\pm}$ 
and thus a difference in the decay widths appears. 
At next-to-leading order we get: 
\begin{equation}
  \G^{\,NLO}\,(H^\pm \to \tau\nu) = 
  \frac{(\mh^2 - m_\tau^2)^2\,y_\tau^2}{16\pi\mh^3}\,
  \left(1\,+\,\frac{2\,\Re\,\d Y_\tau^\pm}{y_\tau}\right)\,.
\label{eq:NLO}
\end{equation}
Here $\d Y_\tau^+$ stands for the decay of $H^+$ and 
$\d Y_\tau^-$ for the decay of $H^-$. 
The radiative corrections  $\d Y_\tau^\pm$ have, in general, 
both CP-invariant and CP-violating contributions:
\begin{equation}
  \d Y_\tau^\pm = \d Y_\tau^{inv} \pm \smaf{1}{2}\,\d Y_\tau^{CP}\,.
  \label{eq:dYi}
\end{equation}
Both the CP-invariant and the CP-violating contributions have real and
imaginary parts.
%
Using eqs.~\eq{NLO} and \eq{dYi}, we can write the CP-violating 
asymmetry $\dcp$ of eq.~\eq{defDCP} in the simple form: 
\begin{equation}
  \dcp =
  \frac{\Re\,\d Y_\tau^{CP}}{y_\tau\,+\,2\,\Re\,\d Y_\tau^{inv}} 
  \simeq \frac{\Re\,\d Y_\tau^{CP}}{y_\tau} \,.
\label{eq:DCP}
\end{equation}

In the MSSM, $\dcp$ gets contributions from loop exchanges of 
charginos, neutralinos, sfermions, $W$ bosons, and neutral Higgs 
bosons. 
The relevant Feynman diagrams are shown in Fig.~1. 
Note that the various diagrams  contribute to $\dcp$ only   
if they have absorptive parts. 
The form factors $\d Y_\tau^{CP}$ can be obtained from $\d Y_b^{CP}$ 
in \cite{Christova:2002ke} by the replacements (s)bottom $\to$ 
(s)tau and (s)top $\to$ (s)neutrino. 
Since $m_\nu \sim h_\nu \sim 0$, many terms vanish and $\d 
Y_\tau^{CP}$
becomes much simpler than $\d Y_b^{CP}$. 
The explicit expressions 
for $\d Y_\tau^{CP}$ due to the various diagrams of Fig.~1, 
together with the masses and couplings of staus and sneutrinos, 
are given in the Appendix.  
All other necessary formulae can be found in \cite{Christova:2002ke}.

\begin{figure}[ht]
{\setlength{\unitlength}{1mm}
\begin{center}
\begin{picture}(151,60)
\put(-13,-164){\mbox{\epsfig{figure=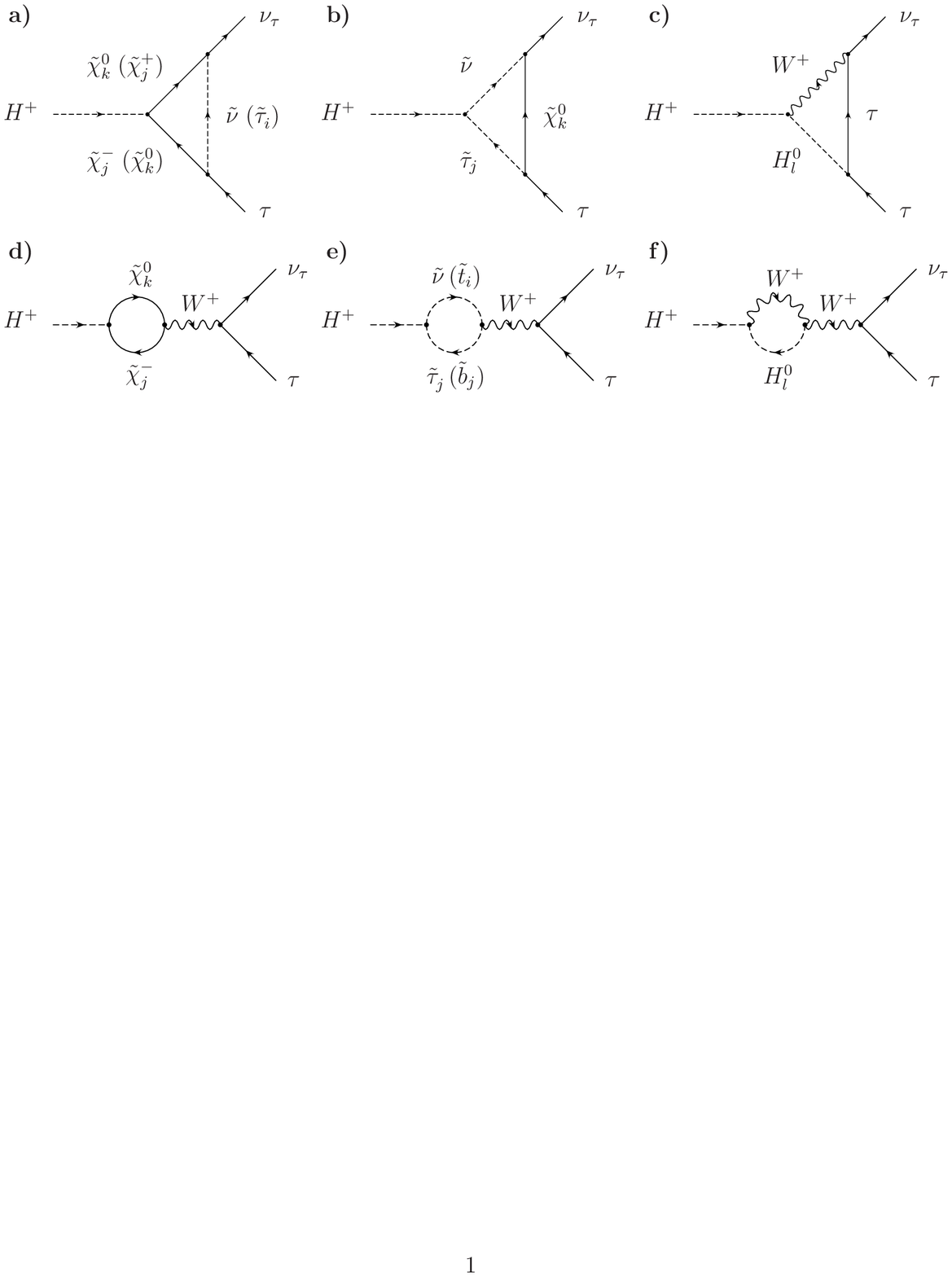,width=17.5cm}}}
\end{picture}
\end{center}}
\caption{Diagrams contributing to CP violation in $H^+\to\tau^+\nu_\tau$ 
  in the MSSM with complex couplings ($j=1,2;$ $k=1,...,4$; $l=1,2,3$).
\label{fig:feyngraphs}}
\end{figure}

\section{Numerical results}

The parameters relevant to our study are 
the gaugino mass parameters $M_1$ and $M_2$; 
the higgsino mass parameter $\mu$;
the soft-breaking parameters of the tau-sleptons
$M_{\ti L}$, $M_{\ti E}$ and $A_\tau$; 
the charged Higgs boson mass $\mhp$, 
and $\tan\beta=v_2/v_1$.
We also need the parameters of the stop-sbottom sector, 
$M_{\ti Q,\ti U,\ti D}$, and $A_{t,b}$: on one hand 
for the $\ti t\ti b$ self-energy diagram of Fig~1e, 
on the other hand for the radiative corrections to 
the neutral Higgs sector. 

Quite generally, the gaugino and higgsino mass parameters 
$M_1$, $M_2$, $\mu$  
and the trilinear couplings of the third generation $A_{t,b,\tau}$ 
can have physical phases 
that may lead to sizable CP-violating effects.  
In order not to vary too many parameters, we fix part of 
the parameter space by 
\begin{align}
   & M_2=200~{\rm GeV},~ \mu=300~{\rm GeV},~
     M_{\ti E} = M_{\ti L} - 5~{\rm GeV},~
     |A_\tau| = 400~{\rm GeV},\label{leppars}\\
   & M_{\ti Q} = 500~{\rm GeV},~
     M_{\ti U} = 450~{\rm GeV},~ 
     M_{\ti D} = 550~{\rm GeV},~
     A_t = A_b = -500~{\rm GeV}.
\label{eq:parset}
\end{align}
For $M_1$, we assume $|M_1|=\frac{5}{3}\tan\t_W\,|M_2|$,  
keeping $\phi_1$, the phase $M_1$, as a physical phase.
The phase of $M_2$ can be rotated away. 
Since according to the measurements of the electron and neutron 
electric dipole moments we have $\phi_\mu < {\cal O}(10^{-2})$ 
\cite{phimu} for SUSY masses of the order of a few hundred GeV, 
we take $\phi_\mu=0$. 
The remaining phases in our analysis are thus 
$\phi_t$, $\phi_b$ and $\phi_\tau$ (the phases of $A_t$, $A_b$
and $A_\tau$) and $\phi_1 $.
These phases also induce, at one-loop level, a mixing of
the CP-even and CP-odd neutral Higgs boson states to form 
mass eigenstates $H_i^0$, i=1,2,3~\cite{Pilaftsis}. 
We take this mixing into account using \cite{CEPW00}.

\begin{figure}[t!]
\setlength{\unitlength}{1mm}
\begin{center}
\begin{picture}(65,65)
\put(0,2){\mbox{\epsfig{figure=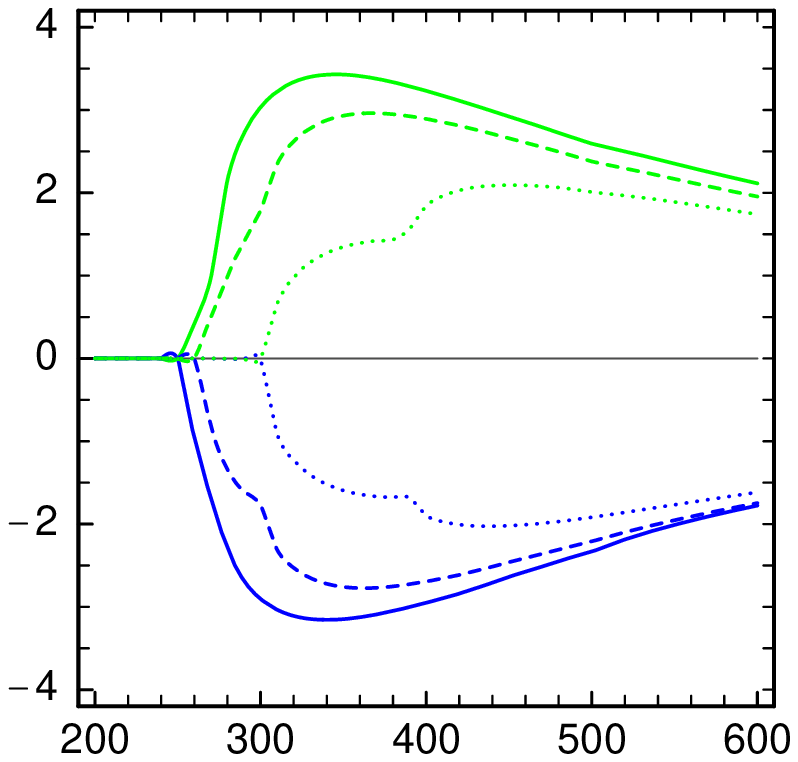,height=65mm}}}
\put(25,-2){$\mh$~[GeV]}
\put(-5,29){\rotatebox{90}{$\dcp~[10^{-3}]$}}
\put(45,16){\footnotesize $\phi_\tau=\pi/2$}
\put(45,58){\footnotesize $\phi_1=\pi/2$}
\end{picture}
\caption{$\d^{CP}$ as a function of $\mhp$
  for $\phi_\tau=\pi/2$, $\phi_1=0$ ($\d^{CP}<0$), and 
  for $\phi_1=\pi/2$, $\phi_\tau=0$ ($\d^{CP}>0$). 
  The full, dashed, and dotted lines are for 
  $\tan\b=5$, 10, and 30, respectively.
  \label{fig:dCPmh}}
\end{center}
\end{figure}

\Fig{dCPmh} shows $\dcp$ as a function of $\mh$ for the two cases 
$\phi_\tau=\pi/2$, $\phi_1 = 0$ and $\phi_\tau=0$, $\phi_1 = \pi/2$
(all other phases zero) 
and three values of $\tan\b$: $\tan\b=5$, 10, and 30. 
$M_{\ti L}$ is chosen such that the lighter stau mass is 
$\mstau{1}=135$~GeV. The corresponding values for $\msnu$, 
$\mstau{2}$ and $\t_{\stau}$ are listed in Table~2.

\begin{table}[h!]\center
\begin{tabular}{|c|r|r|r|r|}
\hline
   $\tan\beta$ & $M_{\ti L}$ & $\msnu$ & $\mstau{2}$ & $\t_{\stau}$ \\
\hline
   5 & 138 & 123 & 150 & $56^o$ \\
  10 & 147 & 132 & 166 & $50^o$ \\
  30 & 180 & 168 & 221 & $47^o$ \\
\hline
\end{tabular}
\caption{Parameters used in the analysis, masses in [GeV], 
$\mstau{1}=135$~GeV.}
\end{table}

The dominant source of CP violation in $H^\pm \to \tau^\pm\,\,\nutau$ 
decays is the sneutrino-stau-neutralino loop of Fig.~1b: 
For $\mh < \mstau{1}+\msnu$, $\dcp$ is negligibly small, 
while it sharply rises once the $H^\pm\to \stau_1^{}\snu_\tau$ channel 
is open.  
In \fig{dCPmh}, $|\dcp|$ goes up to $\sim 3.5\times 10^{-3}$;     
in our analysis, we have not found values larger than 0.5\% 
(though we do not exclude them for some extreme values of MSSM 
parameters). 
Here note that we have taken a rather large value for $|A_\tau|$ 
compared to $M_{\ti L}$. For smaller $|A_\tau|$, $\dcp$ typically 
decreases.  
$\dcp$ also decreases with increasing $\tan\b$.

\begin{figure}[t!]
\setlength{\unitlength}{1mm}
\begin{center}
\begin{picture}(65,65)
\put(0,2){\mbox{\epsfig{figure=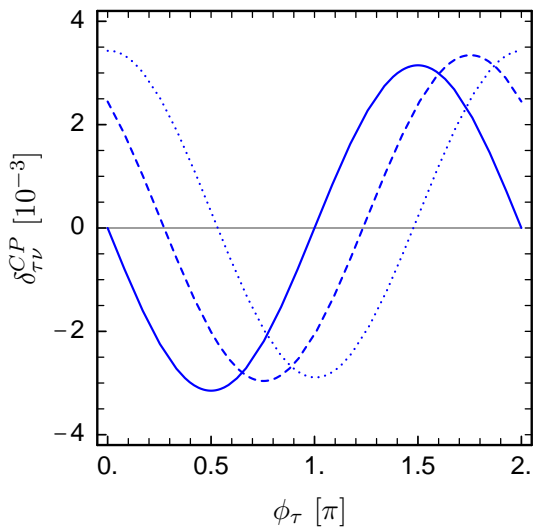,height=65mm}}}
\put(30,-2){$\phi_\tau~[\pi]$}
\put(-5,29){\rotatebox{90}{$\dcp~[10^{-3}]$}}
\end{picture}
\caption{$\d^{CP}$ as a function of $\phi_\tau$ for 
  $\mh=350$~GeV and $\tan\b=5$. 
  The full, dashed, and dotted lines are for 
  $\phi_1=0$, $\pi/4$, and $\pi/2$, respectively.
  \label{fig:dCPphA}}
\end{center}
\end{figure}

It is interesting to note that maximal $\phi_\tau$ and 
maximal $\phi_1$ lead to very similar values of $\dcp$ 
but with opposite signs. 
However, if both phases are maximal, i.e. 
$\phi_t\sim\phi_1\sim\pi/2$ or $3\pi/2$, they compensate 
each other and $\dcp$ practically vanishes. 
In \fig{dCPphA}, $\dcp$ is shown as a function of $\phi_\tau$
for $\mh=350$~GeV, $\tan\b=5$, and 
various values of $\phi_1$.
One sees that $\phi_\tau$ and $\phi_1$ are of equal importance 
for $\dcp$.

We have also examined the dependence on $\phi_{t,b}$.  
Here notice that in the considered range of $\mh$, 
$200~\gev < \mh < 600~\gev$, 
the diagram with $\tilde t\,\tilde b$, Fig.~1e, does not contribute. 
Thus the parameters of the squark sector, eq.~(\ref{eq:parset}), enter 
only through radiative corrections to the neutral Higgs sector,  
see diagrams 1c and 1f, which turn out to be negligible in the case 
$\mh > \mstau{1}+\msnu$.  
For completeness we note that also a non-zero $\phi_\mu$ has 
only little influence on $\dcp$.

\section{Conclusions}

We have calculated the one-loop contributions to the decays 
$H^\pm\to\tau^\pm\,\,\nutau$ within the MSSM with complex parameters.  
They lead to a CP-violating asymmetry $\dcp$, eq.~(\ref{eq:defDCP}),
different from zero. 
The relevant phases in our analysis are those of the 
trilinear coupling $A_{\tau}$ and the gaugino mass $M_1$. 
For $m_{H^{+}} > \mstau{1} + \msnu$, the asymmetry is typically 
of order $10^{-3}$, the dominant source being the 
sneutrino--stau--neutralino loop. 

Some comments are in order on the feasibility of measuring such an
asymmetry.  As already mentioned, the branching ratio for 
$H^\pm \to \tau^\pm\,\,\nutau$ is sizeable only for $\tan\beta > 10$.  
At the LHC, the dominant
production channel for $H^{+}$ is $gb \to H^{+} t$.  One expects
\cite{Belyaev et al.} 1560 events for a Higgs mass of $\mh = 400$~GeV 
and $\tan\beta = 50$ with a ratio signal over background 
$S/\sqrt{B} = 19.8$ with an integrated luminosity of 100 \fb. 
(Here several cuts were already applied and $b$ tagging assumed.) 
With a branching ratio of 22\% one then has 343 events of 
$H^+ \to \tau^+\nu_\tau$.

At a linear $e^{+}e^{-}$ collider at $\sqrt{s} = 1$~TeV, the cross 
section of $e^{+}e^{-} \to H^{+} H^{-}$ for $\mh=400~\gev$ is 6.5~fb. 
Assuming a luminosity of ${\cal{L}} = 500$~\fb ~and again a branching 
ratio of 22\% for $H^{+} \to \tau^+\nu_\tau$ at $\tan\beta = 50$, one 
gets 715 events. 
At CLIC with $\sqrt{s} = 3$~TeV the cross section for $e^{+} e^{-} \to 
H^{+} H^{-}$ for $\mh = 400$~GeV is 3 fb. 
With ${\cal{L}} = 800$ \fb ~one gets 528 $H^{+} \to \tau^+\nu_\tau$ events. 
Therefore, in all cases a higher luminosity would be necessary to observe 
the CP--violating asymmetry $\dcp$.

\section*{Acknowledgements}

This work was supported by the ``Fonds zur F\"orderung der 
Wissenschaftlichen Forschung of Austria'', project no. P13139-PHY, 
and by the EU TMR Network Contract no. HPRN-CT-2000-00-149. 

E.\;C. acknowledges the hospitality and financial support of
the CERN Theory Division. Her work was also supported in part 
by the Bulgarian National Science Foundation, Grant Ph--1010.

\clearpage

\begin{appendix}

\section{Masses and couplings of staus and sneutrinos}

The mass matrix of the staus
in the basis $(\tilde \tau_L,\,\tilde \tau_R)$,
\begin{equation}
  {\cal M}_{\sb}^2 = \left(
  \begin{array}{cc}
    M_{\ti L}^2 - m_Z^2\cos 2\beta\,(\sfrac{1}{2}-\sin^2\theta_W)
    + m_\tau^2 & (A_\tau^*-\mu\tan\beta)\,m_\tau \\
    (A_\tau-\mu^*\tan\beta)\,m_\tau
    & M_{\ti E}^2 - m_Z^2\cos 2\beta\sin^2\theta_W + m_\tau^2
  \end{array}\right) \,.
\end{equation}
It is diagonalized by a rotation matrix $R^{\,\sb}$, 
\begin{equation}
  R^{\,\sb} 
  = \left(\begin{array}{rr}
       e^{\frac{i}{2} \varphi_{\sb}} \cos\theta_{\sb}
    & -e^{\frac{i}{2} \varphi_{\sb}} \sin\theta_{\sb} \\
      e^{-\frac{i}{2} \varphi_{\sb}} \sin\theta_{\sb}
    & e^{-\frac{i}{2} \varphi_{\sb}} \cos\theta_{\sb}
    \end{array}\right) \;,
\end{equation}
such that $R^{\,\sb\,\dagger}{\cal M}_{\sb}^2\, R^{\,\sb} =
{\rm diag}(\msb{1}^2,\,\msb{2}^2)$ and 
${\scriptsize\Big(\!\!\begin{array}{cc} \sb_L \\ 
                                        \sb_R \end{array}\!\!\Big)}
 = R^{\,\sb} \,
 {\scriptsize\Big(\!\!\begin{array}{cc} \sb_1 \\
                                        \sb_2 \end{array}\!\!\Big)}$.
The mass of the left-sneutrino is given by 
\begin{equation}
   \msnu^2 = M_{\ti L}^2 + \sfrac{1}{2}\,m_Z^2\cos 2\beta\,.
\end{equation}
If neutrinos have non-zero masses there is also a right-sneutrino. 
However, since it is electrically neutral and $h_\nu\sim 0$, 
it does not take part in the phenomenology discussed here. 
We thus neglect this state and only consider $\snu\equiv (\snu)_L^{}$.

In the following, we give the Lagrangian for the interactions 
of (s)taus and (s)neutrinos. The other necessary parts of 
the interaction Lagrangian are given in \cite{Christova:2002ke}.
We start with the interaction of Higgs bosons with leptons and 
sleptons:
\begin{align}
{\cal L}_{H\ell\ell} &=
    H^+\bar{\nu}\,(y_\tau^{}\PR)\,\tau^- 
  + H^-\tau^+\,(y_\tau^{}\PL)\,\nu 
  + H^0_l\,\tau^+(s_l^{\tau,R}\PR + s_l^{\tau,L}\PL)\,\tau^- \,,\\
{\cal L}_{H\sq\sq} &=
  G_{\!4j}^{\,\ti\tau}\,H^+\,\ti\nu_\tau^*\,\stau_j +
  G_{\!4j}^{\,\ti\tau\,*}\, H^-\,\stau_j^*\,\snu \,,
\end{align}
with $j=1,2$, $l=1,2,3$ and
\begin{equation}
   \PL = \smaf{1}{2}(1-\g_5)\,, \quad
   \PR = \smaf{1}{2}(1+\g_5)\,.
\end{equation}
For the Higgs boson couplings to leptons we have
\begin{equation}
  y_\tau = h_\tau\sin\b \,, \quad 
  h_\tau = \frac{g\,m_\tau}{\sqrt 2\,m_W\cos\beta}\,,
\end{equation}
and
\begin{eqnarray}
  s_l^{\tau,R} & = &  - \frac{g\,m_\tau}{2\, m_W}\,
    \left(g_{H_l\tau\tau}^S + i\, g_{H_l\tau\tau}^P\right) \,,
    \label{eq:slbRdef}\\
  s_l^{\tau,L} & = &  - \frac{g\,m_\tau}{2\, m_W}\,
    \left(g_{H_l\tau\tau}^S - i\, g_{H_l\tau\tau}^P\right) \,,
  \label{eq:slbLdef}
\end{eqnarray}
The $H^\pm$ couplings to stau and sneutrino are given by 
\begin{equation}
   G_4^{\,\ti\tau} = 
   \left(\! \begin{array}{cc}
      h_\tau m_\tau \sin\b - \rzw\,g\,m_W\sin\b\cos\b \\
      h_\tau\,(A_\tau^*\sin\b + \mu\cos\b) 
   \end{array}\! \right) \; R^{\,\sb}\,.
\label{eq:G4sq}
\end{equation}

\hphantom{.}

\noi
The interactions with charginos and neutralinos are described by
\begin{eqnarray}
  {\cal L}_{\ell\sq\ti\x^+}
  &=& \bar\nu_\tau\,(l_{ij}^{\sb}\PR + 
k_{ij}^{\sb}\PL)\,\ti\x^{+}_j\,\sb_i^{}
    + \bar\tau\,(l_{j}^{\st}\PR + 
k_{j}^{\st}\PL)\,\ti\x^{+c}_j\,\st_i^{}
      \nn \\
  & &  
+\;\overline{\ti\x^{+}_j}\,(l_{ij}^{\sb*}\PL+k_{ij}^{\sb*}\PR)\,\nu_\tau\,\sb_i^*  
+\overline{\ti\x^{+c}_j}\,(l_{j}^{\st*}\PL+k_{j}^{\st*}\PR)\,\tau\,\ti\nu_\tau^*  
\,,\\
  {\cal L}_{\ell\sq\ti\x^0}
  &=& \bar\tau\,(a_{ik}^{\sb}\PR + b_{ik}^{\sb}\PL)\,\nt_k\,\sb_i^{}
   +\bar{\ti\x}^0_k (a_{ik}^{\sb*}\PL + 
b_{ik}^{\sb*}\PR)\,\tau\,\sb_i^* \nn\\
  & & +\;\bar\nu_\tau\,(a_{k}^{\st}\PR + b_{k}^{\st}\PL)\,\nt_k\,\snu 
   +\bar{\ti\x}^0_k (a_{k}^{\st*}\PL + 
b_{k}^{\st*}\PR)\,\nu_\tau\,\ti\nu_\tau^*\,,
\end{eqnarray}
with $i,j=1,2$ and $k=1,...,4$.
The chargino--slepton--lepton couplings are
\begin{align}
  l_{j}^{\st}  &= -g\,V_{j1} \,, &
  k_{j}^{\st}  &= h_\tau\,U_{j2}^* \,, \\
  l_{ij}^{\sb} &= -g\,U_{j1}\Rsb_{1i} + h_\tau\,U_{j2}\Rsb_{2i}\,, &
  k_{ij}^{\sb} &= 0 \,.
\end{align}
The neutralino couplings to slepton and lepton are
\begin{align}
  a_{k}^{\,\st} &= \sfrac{g}{\sqrt 2}\,(\tan\theta_W 
N_{k1}-N_{k2})\,, &
  b_{k}^{\,\st} &= 0\,, \\
  a_{ik}^{\,\sb} &= g f_{Lk}^{\sb}\Rsb_{1i} + 
h_{Rk}^{\sb}\Rsb_{2i}\,, &
  b_{ik}^{\,\sb} &= h_{Lk}^{\sb}\Rsb_{1i} + g f_{Rk}^{\sb}\Rsb_{2i}\,,
\end{align}
with
\begin{align}
  f_{Lk}^{\sb} &= \sfrac{1}{\sqrt 2}\,(\tan\theta_W 
N_{k1}+N_{k2})\,,\\
  f_{Rk}^{\sb} &= -\sqrt{2}\,\tan\theta_W N_{k1}^*\,,\\
  h_{Lk}^{\sb} &= -h_\tau\, N_{k3}^* = (h_{Rk}^{\sb})^*\,.
\end{align}

\noi
The interaction with $W$ bosons is given by
\begin{align}
  {\cal L}_{\ell\ell W} & =
  -\smaf{g}{\rzw}\,( W_\mu^+ \bar\nu_\tau\,\g^\mu\PL\,\tau +
                     W_\mu^- \bar\tau\,\g^\mu\PL\,\nu_\tau ) \,,\\
  {\cal L}_{\sq\sq W} & = -i\sfrac{g}{\sqrt{2}} \left[
      \Rsb_{1i}\, W_\mu^+ (\ti\nu_\tau^* \lrd \stau_i^{})
    + \Rsbs_{1i}\, W_\mu^- ( \stau_i^* \lrd \ti\nu_\tau^{})
  \right]\,,
\end{align}
where
\begin{equation}
  A\delr B = A\,(\partial_\mu B) - (\partial_\mu A)\,B \,.
\end{equation}

\section{CP-violating form factors}

\subsection{Vertex graphs}

\subsubsection*{Neutralino--chargino--sneutrino (stau) loop}

The graph of Fig.~1a, with a neutralino, a chargino, and a
sneutrino in the loop, leads to
\begin{align}
  \Re\,\d Y^{CP}_\tau(\nt_k\ch_j\snu) = \; 
  & \frac{1}{8\pi^2}\,\Big\{
      \Im(F_{jk}^{L}a_{k}^{\st}k_{j}^{\st*})\,
      \Im(B_0(\mhp^2,\mnt{k}^2,\mch{j}^2)) \nn\\
  &\hspace*{-34mm}
   +\left[  m_\tau\mnt{k}\,\Im(F_{jk}^{R}a_{k}^{\st}l_{j}^{\st*})
       +\mch{j}\mnt{k}\,\Im(F_{jk}^{R}a_{k}^{\st}k_{j}^{\st*})
          + \mst{i}^2 \,\Im(F_{jk}^{L}a_{k}^{\st}k_{j}^{\st*})
   \right]\,\Im (C_0) \nn\\
  &\hspace*{-34mm}
   +m_\tau \left[ m_\tau\,\Im(F_{jk}^{L}a_{k}^{\st}k_{j}^{\st*}) \,
         + \mnt{k}\,\Im(F_{jk}^{R}a_{k}^{\st}l_{j}^{\st*})
         + \mch{j}\,\Im(F_{jk}^{L}a_{k}^{\st}l_{j}^{\st*})
   \right]\,\Im (C_2) \Big\} \,,
  \label{eq:dYntchst}
\end{align}
with $C_X=C_X(0,\mhp^2,m_\tau^2,\msnu^2,\mnt{k}^2,\mch{j}^2)$, 
$X=0,2$,  
the three-point functions~\cite{pave} in the notation 
of~\cite{Denner}. 
The contribution from the neutralino--chargino--stau loop has exactly
the same structure. Therefore, $\Re\,\d 
Y^{CP}_\tau(\nt_k\ch_j\stau_i^{})$ 
is obtained from Eq.~\eq{dYntchst} by the following substitutions:
for the masses of the loop particles
$\mnt{k} \to \mch{j}$, $\mch{j} \to \mnt{k}$, $\msnu \to \mstau{i}$ 
and for the couplings
$a_{k}^{\st} \to l_{ij}^{\sb}$, $b_{k}^{\st} \to k_{ij}^{\sb}$,
$k_{j}^{\st*} \to b_{ik}^{\sb*}$, and $l_{j}^{\st*} \to 
a_{ik}^{\sb*}$.

\subsubsection*{Sneutrino--stau--neutralino loop}

The sneutrino--stau--neutralino loop of Fig.~1b gives
\begin{equation}
  \Re\,\d Y^{CP}_\tau(\snu\,\sb_j\nt_k) = \frac{1}{8\pi^2}\,\big[ \,
  \mnt{k}\,\Im (G_{\!4j}^{\,\ti\tau} a_{k}^{\st}b_{jk}^{\sb*})\,\Im (C_0)
  - m_\tau\,\Im (G_{\!4j}^{\,\ti\tau} a_{k}^{\st}a_{jk}^{\sb*})\,\Im (C_2)
  \,\big] \,, 
\end{equation}
with $C_X=C_X(0,\mhp^2,m_\tau^2,\mnt{k}^2,\mst{i}^2,\msb{j}^2)$.

\subsubsection*{W boson--neutral Higgs--tau loop}

For the $W$ boson in the loop we use the $\xi=1$ gauge. 
We thus have to add the corresponding graph with a charged ghost, 
i.e. $W^\pm\to G^\pm$ in Fig.~1c. We get:
\begin{align}
  \Re\,\d Y^{CP}_\tau(W H_l\,\tau) = \,
    & -\,\frac{\sqrt{2}\,g^2}{32\pi^2}\, \Big\{ \Im(X_\tau^R)\,
      \big[ (3 m_\tau^2 - 2 m_{H_l}^2)\,\Im(C_0)
            + 2 m_\tau^2\,\Im(C_2) \nn\\
    & \hspace{18mm}
            +\,\Im\big(B_0(m_{H^+}^2, m_W^2, m_{H_l}^2)\big)
            - 2\,\Im\big(B_0(0, m_\tau^2, m_W^2)\big)\big] \nn\\
    & \hspace{18mm}
            +\,m_\tau^2\,\Im(X_\tau^L)\,\Im( 2 C_0 + C_2) \Big\} \,,
    \label{eq:dY1CP5} 
\end{align}
\begin{equation}
  \Re\,\d Y^{CP}_\tau(G H_l\,\tau) = \,
  -\frac{1}{8\pi^2}\, m_\tau h_\tau\cos\b\,
  \left[ \,\Im\,(\hat X_\tau^R)\,\Im\,(C_0) 
         - \Im\,(\hat X_\tau^L)\,\Im\,(C_2)\,\right] \,,
\end{equation}
where $X_\tau^{R,L} = g_{H_lH^+W^-}^{}\,s_l^{\tau,R,L}$, 
 $\hat X_\tau^{R,L} = g_{H_lH^+G^-}^{}\,s_l^{\tau\,R,L}$, \\
and $C_X=C_X(0,\mhp^2,m_\tau^2,m_\tau^2,m_W^2, m_{H_l}^2)$.

\subsection{Self-energy graphs}

\subsubsection*{Neutralino--chargino loop}

The self-energy graph with a neutralino and a chargino of Fig.~1d 
gives
\begin{align}
  \Re\,\d Y^{CP}_\tau\,(\nt_k\ch_j - W)  =\,
  & \frac{1}{8 \pi^2}\,
    \frac{g^2\, m_\tau}{\sqrt{2}\,m_{H^+}^2\, m_W^2}\,
    \Im\left(B_0(\mhp^2, \mnt{k}^2, \mch{j}^2)\right) 
    \nonumber \\
  & \hspace*{-34mm} \times \, \left[
    \Im\left(c_{II}\right)\mch{j} (\mhp^2 + \mnt{k}^2 - \mch{j}^2) -
    \Im\left(c_{IJ}\right)\mnt{k} (\mhp^2 - \mnt{k}^2 + \mch{j}^2)
    \right]
\label{eq:dY12ntch-W}
\end{align}
with
$c_{II} = F_{jk}^R O_{kj}^R + F_{jk}^L O_{kj}^L\,$, and
$c_{IJ} = F_{jk}^R O_{kj}^L + F_{jk}^L O_{kj}^R$.

\subsubsection*{Sneutrino--stau and stop--sbottom loops}

The graph of Fig.~1e with stau and sneutrino leads to
\begin{align}
  \Re\,\d Y^{CP}_\tau\,(\snu\,\sb_j - W) =\;
  & \frac{g^2}{16\pi^2}\,
    \frac{m_\tau}{\mhp^2 m_W^2}\,(\msb{j}^2-\mst{i}^2) \nn\\
  & \hspace{20mm} \times \,
    \Im\,(G_{\!4j}^{\,\ti\tau} \Rsbs_{1j})\;
    \Im \left(B_0(\mhp^2,\msb{j}^2,\mst{i}^2)\right)\,.
\end{align}
The analogous graph with stop and sbottom gives 
\begin{align}
  \Re\,\d Y^{CP}_\tau\,(\ti t_i\,\ti b_j - W) =\;
  & \frac{3g^2}{16\pi^2}\,
    \frac{m_\tau}{\mhp^2 m_W^2}\,(m_{\ti t_i}^2-m_{\ti b_j}^2) 
    \nn\\
  & \hspace{14mm} \times \,
    \Im\,(G_{\!4ij}^{\,\ti t} R^{\ti t}_{1i} R^{\ti b*}_{1j})\;
    \Im \left(B_0(\mhp^2,m_{\ti b_j}^2,m_{\ti t_i}^2)\right)\,, 
\end{align}
where $G_{\!4ij}^{\,\ti t}$ is the $\ti t\ti b H^+$ coupling, 
see eqs.~(48)--(49) of \cite{Christova:2002ke}.

\subsubsection*{\boldmath $W^\pm$--$H^0_l$ and $G^\pm$--$H^0_l$ loops}

The self-energy graph with $W^+$ and $H^0_l$ is shown in Fig.~1f.
Since we use $\xi=1$ gauge for the $W$ in the loop,
we have to add the corresponding graph with a ghost,
i.e. $W^\pm\to G^\pm$ in the loop.
(The second $W$ propagator can be calculated in the unitary gauge.
Hence, no ghost is necessary in this case.)
The two contributions together give:
\begin{align}
  \Re\,\d Y^{CP}_\tau\,(W H^0_l - W) =\;
  & -\,\frac{1}{32\pi^2}\,
    \frac{g^3\,m_\tau}{\sqrt{2}\,\mhp^2 m_W^{}}\,
    (2m_W^2 - 2m_{H_l}^2 - 3\,\mhp^2) \nn\\
  & \times \,  
    O_{3l}\,(\cos\b\,O_{1l} + \sin\b\,O_{2l})~
    \Im\left(B_0(\mhp^2,m_{H_l}^2,m_W^2)\right)\,.
\end{align}

\end{appendix}



\begin{thebibliography}{99}

\bibitem{Ibrahim-Nath}
For a recent review, see: T. Ibrahim and P. Nath, 
hep--ph/0210251.

\bibitem{Christova:2002ke} 
E. Christova, H. Eberl, S. Kraml and W. Majerotto, 
Nucl.\ Phys.\ B {\bf 639} (2002) 263, 
erratum ibidem to appear, hep-ph/0205227. 

\bibitem{phimu} P. Nath, Phys. Rev. Lett. {\bf 66} (1991) 2565; Y. 
Kizukuri and N. Oshimo, Phys. Rev. {\bf D46} (1992) 3025; R, Garisto 
and J.D. Wells, Phys. Rev. {\bf D 55} (1997) 1611;  Y. Grossman, Y. 
Nir and R. Rattazzi, Adv. Ser. Direct. High Energy Phys. {\bf 15} 
(1998) 755.

\bibitem{Pilaftsis} 
A.~Pilaftsis, 
     Phys.\ Rev.\ D {\bf 58} (1998) 096010 [hep-ph/9803297] and 
     Phys.\ Lett.\ B {\bf 435} (1998) 88 [hep-ph/9805373];
A.~Pilaftsis and C.~E.~Wagner,
     Nucl.\ Phys.\ B {\bf 553} (1999) 3 [hep-ph/9902371];
D.~A.~Demir,
     Phys.\ Rev.\ D {\bf 60} (1999) 055006 [hep-ph/9901389];
T.~Ibrahim and P.~Nath, Phys. Rev. D66 (2002) 015005.

\bibitem{CEPW00} M.~Carena, J.~R.~Ellis, A.~Pilaftsis and 
C.~E.~Wagner,
                 Nucl.\ Phys.\ B {\bf 586} (2000) 92 [hep-ph/0003180];
              the Fortran program {\tt cph.f} can be obtained from 
              {\tt http://pilaftsi.home.cern.ch/pilaftsi/}

\bibitem{Belyaev et al.} 
A. Belyaev, D. Garcia, J. Guasch, J. Sola, JHEP 0206, (2002) 059.

\bibitem{pave} G.~Passarino and M.~J.~Veltman,
               Nucl.\ Phys.\ B {\bf 160} (1979) 151.

\bibitem{Denner}
A.~Denner, Fortschr. Phys. {\bf 41} (1993) 307.


\end{thebibliography}
\end{document}